\documentstyle[aas2pp4,psfig]{article}
\received{}
\accepted{}
\journalid{}{}

\newcommand{\ha}{H$\alpha$}
\newcommand{\msun}{M$_\odot$}
\newcommand{\zsun}{Z$_\odot$}

\slugcomment{}
\lefthead{Korchagin, Mayya, Vorobyov}
\righthead{}
\newcommand{\be}{\begin{equation}}
\newcommand{\ee}{\end{equation}}
\begin{document}

\title{Optical Color Gradients in Star-Forming Ring Galaxies}
\author{V. Korchagin$^{1}$, Y.D. Mayya$^2$ and E.I. Vorobyov$^1$}
\affil{$^1$Institute of Physics, Stachki 194, Rostov-on-Don, Russia\\
Email: vik@rsuss1.rnd.runnet.ru}
\affil{$^2$Instituto Nacional de Astrofisica, Optica y Electronica, Apdo
Postal 51 y 216, C.P. 72000, Puebla, M\'exico \\
Email: ydm@inaoep.mx}
\vskip 10pt

16 pages, 12 figures, Accepted for publication in 
the Astrophysical Journal 
 
\begin{abstract}
We compute radial color gradients produced by an outwardly propagating
circular wave of star formation and 
compare our results with color gradients observed in the classical
ring galaxy, the ``Cartwheel''. We invoke two independent models
of star formation in the ring galaxies. The first one is the
conventional density wave scenario, in which an intruder galaxy
creates a radially propagating density wave accompanied by
an enhanced star formation following the Schmidt's law. The second 
scenario is a pure self-propagating star formation model, in which 
the intruder only sets off the first burst of stars at the point of impact.
Both models give essentially the same results.
Systematic reddening of $B-V$, $V-K$ colors towards the center,
such as that observed in the Cartwheel, can be obtained only if the 
abundance of heavy elements in the star-forming gas is a few times below solar.
The $B-V$ and $V-K$ color gradients observed in the Cartwheel can be explained
as a result of mixing of stellar populations born in a star-forming wave
propagating through a low-metallicity gaseous disk, and a pre-existing 
stellar disk of the size of the gaseous disk with color properties typical 
to those observed in nearby disk galaxies.  

\end{abstract}
%
%
      \section{INTRODUCTION}
%

Ring galaxies are believed to be the result of a head-on galaxy-galaxy 
collision.  Such a collision generates an outwardly 
propagating wave of density enhancement in the disk of the larger
galaxy, referred to as the target. Numerical models, starting from the 
pioneering paper of Lynds \& Toomre (1976), confirm this picture. 
Simulations reproduce some of the basic features observed in ring galaxies, 
such as rings or crescent-shaped  structures. 

The density enhancements are likely to trigger star formation, and 
a ring of star formation is expected to form as a response to a circular 
density wave. The star forming ring moves with time to progressively outer
radii as the density wave propagates outwards,
and leaves behind an evolved stellar 
population with the youngest stars located at the current position of the wave.
The star formation history in a propagating wave is preserved, therefore, 
in the radial color distributions of stars in the disk.
The inner disk is hence expected to be redder than its outer parts.
Colors systematically reddening towards the center were indeed observed
in the Cartwheel galaxy (Marcum, Appleton \& Higdon 1992, hereafter MAH), 
thus confirming this basic picture. 
More recently, similar radial color gradients were also found in other ring 
galaxies (Appleton \& Marston 1997).

Although there seems to be a good qualitative agreement between the 
collisional density wave model and the observed color gradients in ring 
galaxies, there is not yet a model of star formation that can
quantitatively explain the observed color gradients.
In their attempt to explain the color gradients in the Cartwheel,
MAH had assumed a constant star formation scenario (Struck-Marcell \& 
Tinsley 1978), and an instantaneous burst of star formation 
(Charlot \& Bruzual 1991). 
The agreement between these models and the data is far from satisfactory.
Both the models fail to reproduce the observed sequence of the 
reddening in the Cartwheel's disk. 
Part of this mismatch is
likely due to a lack of consistency between the data and the models
they used, as discussed below.
MAH measured the color gradients in the Cartwheel 
in nine annuli with the width $\approx 1.3$  kpc each. A wave propagating
with the velocity of 60--90 km\,s$^{-1}$ takes $1.4-2.2 \times 10^7$ years
to cross an annulus. This is clearly inconsistent with the 
assumption of the instantaneous burst model that all the stars in an
annulus were born simultaneously. The problem is more serious if the 
wave propagates more slowly. 
Constant star formation models assume an extended duration of star
formation at each ring, whereas in an expanding density wave the
star formation stops once the wave sweeps across that radius. 
Solar metallicity had been used in both the models,
whereas the Cartwheel's metallicity is about a tenth of solar 
(Fosbury \& Hawarden 1977). 
The metallicity is one of the basic parameters which controls stellar
evolution (Schaller et al. 1992) and hence color gradients are expected 
to depend on the metallicity.

In this paper we re-address the theoretical modeling of radial color
gradients in ring galaxies taking into account the age-spread within
a sampling annulus, and the metallicity effects. 
We invoke two independent scenarios of star formation in ring galaxies.
The first one is the conventional density wave scenario.
In this scenario, the intruder's passage through the
disk of the target galaxy creates a density wave which
compresses the gas, thus inducing star formation. Star formation rate at
each radius in this model is basically controlled by the strength of the
density wave at that radius and the gas density. We use the Schmidt's
law (Schmidt 1959) to relate star formation with the gas density.
In the second scenario, we invoke a pure self-propagating star formation model,
in which star formation is dissociated from the effects of
the density wave model. 
An intruder has only a role of triggering the first event of
star formation close to the point of impact; phenomena related to massive
stars control the star formation at all other radii. 
As a particular case of the latter scenario  
we use the ``fire in the forest'' (hereafter FIF) model earlier applied
for the interpretation of the surface brightness and the chemical
abundance gradients in ring galaxies (Korchagin et al. 1998; Korchagin
et al. 1999).

The kinematical data of ring galaxies seem to favor the density wave
model for ring galaxies. However, the density wave model has a 
shortcoming in the case of the Cartwheel ---
none of the companion galaxies of the Cartwheel have enough mass to 
generate density waves strong enough to form a giant ring such as
that observed in the Cartwheel (Davies \& Morton 1982). New mass estimates
have not improved the situation (Higdon 1996).  
Under the FIF scenario, star formation can be 
sustained subsequently even if the density wave is weak or totally absent.
FIF model completely neglects, however, any dynamical effects,
which is a serious shortcoming of the theory. Both models, however, suit
equally well to our purpose to interpret optical color gradient
in a radially propagating star forming wave. 

Although we treated the two scenarios as completely independent
in this work, the two mechanisms may operate simultaneously.
FIF phenomena can play a role even when density wave by itself
is strong enough to trigger star formation.
The real situation may be a combination of the two scenarios.

In Section 2, we formulate the models of star formation in ring galaxies.
In Section 3, we use these models to compute color-color diagrams for stellar 
populations born in the wave of star formation, propagating in a purely 
gaseous disk with different metallicities. In Section 4, we discuss the 
comparison of the theoretical radial color gradients with the observed 
radial color gradients in the disk of the Cartwheel.\
The results are discussed 
in Section 5. 

%
	\section{STAR FORMATION SCENARIOS IN RING GALAXIES}
%

\subsection{Density Wave Induced Star Formation}

Classical models of ring formation through collisions between galaxies predict
an outwardly propagating circular density wave in the disk of the target
galaxy. The density wave is expected to compress the gas, thus
increasing the gas densities above the critical values required for
star formation. We assume the star formation rate is proportional
to the square of the density enhancement $ \tilde C(r-Vt)$ propagating outwardly
in a homogeneous gaseous disk.
The balance of the surface density of stars ($M_s$) at a distance $r$
from the center propagating with a velocity $V$ and at time $t$ can be 
then written as follows:
 
\be
    {dM_s\over dt} = -D + k b {\tilde C}^2(r-Vt),
\label{eq1}
\ee
with the density enhancement ${\tilde C}$ given by a Gaussian
function:
\be
   {\tilde C}(r-Vt) = A\exp{\Big[-{(r-Vt)^2 \over l^2} \Big]}.
\label{eq2}
\ee
The first term $D$ in the right-hand side of equation (1)
gives the decrease of stellar density due to the death of stars, and the 
second term describes the growth of density of stars as a result of
star formation. The coefficients $b$ and  $k<1$ determine the
rate of star formation and its efficiency. 
$l$ is the width of a star forming ring.

Any dependence of the density enhancement on the strength of the density wave
is not included in the above expression.
Generally, the circularly expanding density waves are expected to change
their amplitude with time. Numerical simulations by Athanassoula et al. (1997)
and Struck-Marcell \& Lotan (1990) show
that the relative amplitude of the density wave remains near-constant
during wave propagation.
The amplitude of the density wave approximately follows the
surface density profile of an unperturbed disk which
justifies our assumption of constant density wave amplitude
for the  flat surface density profile used in Section 3.
If the pre-collision Cartwheel was a normal
disk galaxy with an exponential surface density distribution, 
the amplitude of the density wave should
decrease more than an order of value when the wave reaches the
present position of the outer ring of the Cartwheel.

\subsection{FIF Wave of Star Formation}

The propagation of FIF star formation wave in a gaseous medium with an
initial surface mass density, $M_c$
is described by the set of equations (Korchagin et al. 1998):
\be
   {dM_s\over dt} = - D + k a M_c({\bf x},t-T) \int d{\bf x}^\prime
f({\bf x}
-
                     {\bf x}^\prime)M_s({\bf x}^\prime , t - T),
                                         \label{eq3}
\ee
\be
    {dM_c\over dt} = - aM_c({\bf x},t) \int d{\bf x}^\prime f({\bf x} -
                     {\bf x}^\prime)M_s({\bf x}^\prime , t).
                                          \label{eq4}
\ee
Equations (3) and (4) give the rate of increase of surface density of
the stars  $M_s$, and the corresponding decrease of surface density
of the star forming gas $M_c$ respectively. The first term $D$ in the 
right-hand side of equation (3) describes the decrease of stellar
density due to the death of stars, and the second term describes the
growth of density of stars as a result of
induced star formation. The coefficients $a$ and $k<1$
determine the rate of star formation and its efficiency. 
The function $f$ models the nonlocal ``influence''
of the star complexes located at point $\bf x^\prime$ on the interstellar
medium at point $\bf x$. This function has a characteristic  ``radius of 
influence'' $L$, which, together with the characteristic time of the 
formation of stars $T$, controls the velocity of the star-forming wave. 

\subsection{Model Parameters}

For the density wave scenario, computations were carried out
for propagation velocities of 25, 55, 90 and 120 km\,s$^{-1}$. The efficiency of
star formation $k$ was taken to be 0.1. The amplitude
$A$ of the density wave was chosen to be $A = 1.5 \times 10^7$ \msun/kpc$^2$.

The set of parameters, as well as the particular choice of the function $f$
for the FIF scenario, were taken from Korchagin et al. (1998). 
Specifically, $L = 2$~kpc, the delay parameter $T=1.8\times10^7$ yr, the
efficiency of star formation $k = 0.1$.
We refer the reader to Korchagin et al. (1998) for more details. 

The parameters $a$ and $b$ determining the rate of star formation
were chosen so as to reproduce
the observed \ha\  surface brightness profile in the Cartwheel (Higdon 1995).
This gives a value of $a = 0.75$, and $b = 0.025$ when masses, time scales
and distance scales are expressed in units $10^7$\msun, $10^6$ yr
and 1 kpc respectively. With such values of parameters $a$ and $b$,
the theoretical rate of star formation in the wave is about 
$1.6 \times 10^{-7}$ \msun yr$^{-1}$ pc$^{-2}$ which is comparable
to the estimated rate of star formation in the Cartwheel 
$\sim 2 \times 10^{-7}$ \msun yr$^{-1}$ pc$^{-2}$ (Higdon 1995).
Note that the colors of the star-forming waves
are luminosity ratios by definition and
do not depend on the free parameters controlling the rate of star formation.

The pre-collisional gas density is assumed to be uniform
for purposes of studying the general color properties of 
the star-forming waves. 
In Section 4.3, 
we discuss the effects of wave propagation in a medium with an exponentially
decreasing density profile.
The colors
of the star forming wave propagating in a medium with exponential
surface density profile do not differ significantly from those
computed for the flat surface density profiles.


\section{COLOR GRADIENTS IN A STAR-FORMING WAVE } 


\subsection{Population Synthesis Model}

The mass $M_s$ of stars formed per unit area is distributed into 
individual stellar masses using the Salpeter's IMF with $\alpha = 1.35$, 
and the stellar mass interval of $0.1~{\rm M}_\odot \leq m_s \leq
100~{\rm M}_\odot$. 
Once the IMF is fixed, the birth and death rates of stars can be 
determined at each radial grid zone from equations (1)--(4).
The luminosity at a given band $A$ after time $t$ can be computed as :
\be
L_A(t) =   \sum\limits_{m}\sum
\limits_{\tau} l_A(m,\tau) N(m,\tau,t),
\label{eq5}
\ee
where $N(m,\tau,t)$ is the number of stars per unit surface area at time $t$
with mass $m$ and age $\tau$, and $l_A(m,\tau)$ is the luminosity of a 
star in a band $A$. Stellar luminosities $l_A(m,\tau)$ are obtained 
using stellar evolutionary (Schaller et al. 1992) and 
atmospheric (Kurucz 1992) models. 
To determine the dependence of the color gradients on
the chemical abundances in the star-forming gas,  
the initial metallicities adopted ranged from Z$_{\odot}/20$ to  2Z$_{\odot}$.
The population synthesis technique used in this work is explained in
detail in Mayya (1995). The results of this code are compared with those
from other existing codes by Charlot (1996). Recent updates in the code
are described in Mayya~(1997). 

To study general color properties of stellar populations produced by the
waves of star formation, we modeled the propagation of the density wave in
a purely gaseous homogeneous disk. To do this, we divided the disk
into thirteen annuli with widths 1.25 kpc each. The first four annuli
represent the  ``nucleus'', and the outer annuli correspond
approximately to the positions of annuli in Figure 1a of MAH.
The approach was used both for the density wave, and
FIF-wave models.
The Equations (1)-(5) were solved numerically, and
the average colors were computed in each annulus and in the ``nucleus'' for
the different velocities of the wave propagation, and for the different 
metallicities of the star-forming gas. Specifically, we computed color gradients 
produced by the wave, propagating with the velocities 25, 55, 90 and 
120 km\,s$^{-1}$, and for the gas metallicities equal to Z$_{\odot}/20$,
Z$_{\odot}/5$, Z$_{\odot}/2.5$, Z$_{\odot}$, and 2Z$_{\odot}$.

\subsection{Results for Density wave and FIF wave}

Figures 1--4 show the results of these simulations for the density wave model
at four different assumed velocities.
Each frame in
Figures 1--4 presents the color-color diagram computed for the moment when
the star-forming wave reaches 16 kpc, the present position of the
outer ring in the Cartwheel. 

Two conclusions can be drawn from the theoretical color diagrams.
Color properties of stellar populations produced by the
wave of star formation do not depend strongly on the velocity of the wave,
but are sensitive to the heavy element abundance of the star-forming gas.
Stellar populations in the wake of the star-forming wave have spatially
sequenced colors similar to those observed in the Cartwheel if the gas
metallicity  is sub-solar.
Figures 1--4 show that color diagrams do not have regular reddening towards the
center of the disk if metallicity of the star-forming gas is solar or higher.
This result is consistent with the measurements of the elemental abundances
in the Cartwheel. In this galaxy, the heavy element abundances are deficient
by a factor of ten as compared to the Orion Nebula (Fosbury \& Hawarden 1977).

The origin of color sequentiality in a low-metallicity star-forming wave
lies in the dependence of color-magnitude diagrams upon metallicity.
Lowering Z causes individual stars to become both brighter and hotter.
The relative contribution of supergiants to the total continuum
considerably decreases with decreasing of metallicity.
E.g., the relative contribution of supergiants to the total continuum
in R-, and V-bands in a 5--20 Myr starburst is about 20--80 percent
for solar metallicity, and decreases to 3--40 percent
for $ Z = Z_{\odot}/10$ (Mas-Hesse \& Kunth 1991).
These factors, together with smoothening of colors
by a non-coeval star formation in photometric rings, lead to the
ordering of color-color diagrams in a low-metallicity gas.
At solar metallicities photometric rings containing red supergiants 
break the spatial ordering of colors.
Uncertainties of supergiant models do not allow to make exact predictions,
but general trend of ordering of colors in a low-metallicity
star-forming wave is robust.
 
For comparison purposes, we also computed colors for a FIF scenario, 
in which all the effects of a density wave model are completely excluded. 
Figure 5 presents color diagrams
for the FIF wave propagating with the velocity 90 km\,s$^{-1}$
computed by solving equations (3), (4) and (5). The IMF,
stellar mass intervals, and the chemical abundances of the star-forming gas
were chosen to be the same as in the density wave simulations shown in Figure 3.
Comparison of these two figures further demonstrates that the color
properties of stellar populations born in a star-forming wave do not
depend on the particular mechanism of the wave propagation.
This is because wave velocity and the present radius of the ring 
basically determine the mean age, and hence the color, of the stellar 
populations at a given radius. 

We studied the dependence of color gradients on the particular choice
of the initial mass function. Namely, we calculated color gradients for 
the Salpeter IMF with stellar mass interval of $0.5~{\rm M}_\odot \leq m_s \leq
50~{\rm M}_\odot$. Such an IMF produces more luminous stellar populations
for a given stellar mass. Simulations show that colors are not 
much affected by the choice of the stellar mass intervals of IMF. 
The choice of IMF affects, however,
luminosities of stellar populations born in the wave. 
In general, radial color gradients are strongly affected by abundance of
heavy elements in the star-forming gas, and to less extent are affected by
the radial velocity of wave of star formation. Colors are insensitive
to a particular mechanism of the star-forming wave, and to
the mass intervals of IMF.  

\section{APPLICATION TO THE CARTWHEEL}

Theoretical colors in the wave are bluer compared to the colors observed 
in the Cartwheel galaxy (Fig. 8 of MAH), and hence the Cartwheel's colors
cannot be explained by a wave of star formation propagating in a purely
gaseous disk. This implies that a detailed modeling is required to fully 
understand the color gradients in the Cartwheel.
In this section, we describe the method that we have
adopted to explain the observed colors in the Cartwheel.
 
\subsection{Extinction Correction}

In principle, the redder colors of the Cartwheel in comparison to those
of the expanding wave models can be the result of an increasing dust 
content towards the nucleus. There is evidence for both cold and
warm dust in the Cartwheel. The cold dust is inferred
from the HST images, which show a complicated network of dust lanes in
the inner ring regions. The warm dust is inferred in the center regions
and in part of the outer ring by ISOCAM observations 
(Charmandaris et al. 1999). However, there has not
been an estimation of the extinction gradient from the above studies. The ISOCAM
emission originates from non-equilibrium heating of Very Small Grains
and Polycyclic Aromatic Hydrocarbons which do not contribute much to the
dust mass and hence to the optical extinction (Mayya \& Rengarajan 1997).
Therefore even the high central emission in ISOCAM bands cannot be taken as
evidence for a higher dust mass at the center.

We now investigate whether the observed gas densities in the Cartwheel
are consistent with the assumption that the extinction plays a major role 
in creating an ordered color gradient.
We hence estimated the color excess $E_{B-V}$ based on the hypothesis
that the difference between the observed color gradient and
that for the wave propagating on a purely gaseous disk, can be completely
explained by an extinction gradient. Details of the wave model that we
used for the Cartwheel will be explained in Sec.~4.3. Using the galactic 
dust/gas ratio (Bohlin et al. 1978), we then estimated the 
expected total surface density of gas $\Sigma_{HI+H_{2}}^{ext}$.
The corresponding values of $E_{B-V}$ for each annulus are listed in
column 2 of Table 1, and the estimated surface densities of
gas $\Sigma_{HI+H_{2}}^{ext}$ are given in column 3.
Columns 4, 5, and 6 in Table 1 give the observed surface densities of atomic
hydrogen $\Sigma_{HI}^{obs}$, molecular hydrogen $\Sigma_{H_{2}}^{obs}$,
and the sum of atomic and molecular hydrogen $\Sigma_{HI+H_{2}}^{obs}$
in each annulus (Higdon 1996; Horellou et al. 1998).
The annuli corresponding to the inner ring  
and two outer rings (VIII and IX) have the observed gas densities 
consistent with the extinction-estimated values.
In the rest of the annuli, the extinction-estimated gas densities are
overestimates, implying that the real extinction value is lower than that
required to explain the observed colors.
Hence the extinction gradient cannot be responsible for the observed 
color gradient. 
Additional argument against large amounts of extinction is low
metallicity of gas which decreases dust/gas ratio.
This agrees with de Jong (1996)
who found that the reddening due to dust does not play a significant role
in the explanation of color gradients in the disks of spiral galaxies.
The color gradients in galactic disks result
from the combination of differences in stellar ages and
the metallicity gradients across the disks.

The observed gas densities, on the other hand, can be used to derive
the visual extinction at each annuli. The galactic dust/gas ratio
gives us a value of 0.32 mag at the annuli II--VII and 1.2 mag
at the two outermost annuli.
Using Balmer line ratios, Fosbury \& Hawarden (1977) found the extinction
in two outer HII regions to be $A_v=2^m.1$.
Higdon (1996) derived an upper limit to visual extinction in two outer
HII regions CW 17 and CW 23 of $A_v=1^m.6$ and $A_v=2^m.1$ which are
consistent with the estimate by Fosbury \& Hawarden (1977).
Unfortunately, very little is known about the azimuthally
averaged visual extinction, which is the directly relevant quantity
for determining the corrections to the radial color gradients.
The values reported by Fosbury \& Hawarden (1977) and Higdon (1996) are 
greater than the observed visual extinction in other extragalactic HII regions.
Kaufman et al. (1987) found an averaged  value of $<A_v>=1^m.1 \pm 0.4$ 
for 42 HII regions in M81. The azimuthally averaged visual extinction in 
the Cartwheel's outer ring can also be lower than the observed values 
for two most luminous HII regions. Visual extinction estimates
based on the observed gas densities give also a lower value, which is
closer to the value found by Kaufman et al (1987) than that estimated 
from Balmer lines. Note also that for $A_v = 2^m.1$,
the visual extinction correction to the observed colors in the two
outer rings of the Cartwheel gives unrealistically blue colors. These colors
are bluer than the theoretical colors of a very young (1 Myr) starburst.
We adopt therefore in our calculations, the values $A_v = 1^m.2$ for the
two outermost photometric rings, and $A_v = 0^m.32$ for the rest of the 
photometric rings, following the estimates based on the observed gas densities.

\subsection{The Wave Velocity}

Even though the model colors are not strongly affected by the wave velocity,
we discuss here the velocity adopted because of its importance
for understanding the physical nature of the star-forming waves.
  
There have been studies of kinematical motions in the 
Cartwheel galaxy using different techniques. Fosbury \& Hawarden (1977)
used optical long-slit spectroscopic data, Higdon (1996) used HI data 
and more recently Amram et al. (1998) used \ha\  Fabry-Perot data. 
A standard method of interpretation of motions in the ring galaxies
assumes that all the internal motions are in the plane
of the ring, and that the kinematical motions can be represented by
a systematic motion of the galaxy together with ring's rotation and 
its expansion.
Using this method, different authors obtain rather different values
for the expansion velocity of matter in the outer ring of the Cartwheel galaxy.
Fosbury \& Hawarden (1977) used optical spectroscopy of 5 HII regions and
found a radial expansion velocity of $89\pm40$ km\,s$^{-1}$.  Higdon (1996)
finds that the outer ring is expanding with the velocity $53\pm9$ km\,s$^{-1}$.
Amram et al. (1998) using the best spatial resolution and much larger number of
points compared to the previous studies find the velocity of expansion 
(13--17) $\pm 2$ km\,s$^{-1}$. With a slightly different approach, Amram et al.
find the value of the matter expansion to be $30\pm10$ km\,s$^{-1}$.
This value is in poor agreement with the value $53\pm9$ km\,s$^{-1}$
found by Higdon (1996). The authors notice, however, that cold and 
ionized gas might suffer different expanding phases.
 
It must be stressed that all these measurements determine the
internal motions of cold or ionized gas in the wave of
star formation, and do not determine the velocity of propagation
of the wave itself. This is exactly similar to the kinematical studies 
of spiral density waves, where the measured motions within the spiral arms
themselves (i.e. the streaming motions)
determine the amplitude of spiral density waves, and do not give 
the pattern speed of the wave.
In our modeling, we therefore based the
choice of the propagation velocity of the star-forming wave
on the best fit for the  surface brightness distributions
in the disk of the Cartwheel.
Korchagin et al. (1998) found that
the observed \ha\  and red continuum surface brightness distributions
are well reproduced for the velocity of the wave of about 90 km\,s$^{-1}$.

\subsection{Color Gradients in the Cartwheel}

The color gradients that were presented in the previous section
assumed a homogeneous gaseous disk. The majority of galaxies 
have exponentially decreasing gas surface density profiles.
Nevertheless, the above calculations adequately 
illustrate the basic dependence of color gradients as a function of the wave 
velocity and metallicity. To calculate the color gradients in
the Cartwheel, we used an exponential gas surface density 
profile 
\be
\sigma_{c}(r)=A\exp(-r/H)
\label{eq6}
\ee
with the 
parameters $A=1.5 \times10^8$ \msun\,kpc$^{-2}$ and $H=6.5$\,kpc which are
typical for the late-type spirals (Tacconi \& Young 1986).
The initial metallicity of gas was chosen to be one-twentieth of the solar 
metallicity, close to the values observed for the 
outer ring by Fosbury \&  Hawarden (1977). 

FIF wave propagating in an inhomogeneous medium has a non-constant 
velocity which affects the ages of stellar populations behind the 
star-forming wave, and hence their color properties. The change in the 
wave velocity due to the radial variation of surface density is within 12\% 
of the mean value 94 km\,s$^{-1}$ throughout the disk of the Cartwheel. 
The resulting model color-color diagram is shown in Figure 6. Colors do not 
differ significantly from those computed in the previous section 
(see the top-left plot in Figure 5). Figure 6 also includes
the extinction-corrected colors observed in the Cartwheel (filled triangles).
The extinction corrections were based on the gas column density and the
galactic $A_v/N_{\rm H}$ ratio as explained in the
previous subsection. Specifically, the observed colors for the two outer 
photometric rings were corrected for the extinction $A_v= 1^m.2$, and 
the rest were corrected for the extinction value $A_v = 0^m.32$.
The model reproduces the sequence of points in the color-color diagram. 
However $V-K$ colors in the wave are around 1 magnitude bluer than 
the observed colors of the Cartwheel galaxy. 
Hence, the color properties of the stellar populations born in a purely 
star-forming wave together with the extinction correction cannot 
explain entirely the observed colors. This is not unexpected,  because 
the pre-collisional stellar disk is known to have a non-negligible 
contribution to the colors of Cartwheel. 

Optical color gradients are common among ``normal'' disk galaxies.
Figure 7 shows $B-V$ vs $V-K$ color profiles in the disks of six nearby
galaxies which were plotted using de Jong's (1996) observational data.
The observed colors of spiral galaxies are even redder than the observed
colors in the Cartwheel.
This indicates that the Cartwheel's colors can indeed be obtained by
mixing stellar populations born in the wave with the older stellar populations
existing in the Cartwheel's disk before the collision.

Color mixing cannot be considered separate from the surface brightness
constraints. 
The spatial distribution of the young stellar population 
can be found from the best fit to $H_{\alpha}$ surface 
brightness distribution in the Cartwheel. 
Assuming exponential gas surface density distribution
we find that the best fit to $H_{\alpha}$ surface
brightness distribution in the Cartwheel (Figure 8) can be obtained
for the FIF-wave propagating in the gaseous disk with the
parameters $A = 5\times 10^7$\msun\,kpc$^{-2}$, and $H=11 kpc$.
For the old stellar population in the Cartwheel we assumed that 
it has an exponential  distribution in the $V$-band.  
Specifically,
we assumed that V-band surface brightness $\mu(r)$ at a
distance $r$ from the
galactic center is given by the relation
$\mu(r) = \mu_0 + 1.086r/R_0$,
where $\mu_0$ is the surface brightness in the center of the disk
in mag\,arcsec$^{-2}$, and $R_0$
determines the scale length of the exponential surface brightness distribution.
Choosing $R_0=5$ kpc which is the typical value for disk galaxies (de Jong 1996)
we find that the best fit to the red continuum profile in the
Cartwheel can be obtained for 
$\mu_0=23.5$~mag\,arcsec$^{-2}$. 
The red continuum brightness distribution
is however in a poor agreement with observations.
Maximum of the red continuum peak associated with the outer ring
of the Cartwheel is less prominent as compared to observations.
Higher gas metallicities
increase a relative contribution of supergiants to the total
continuum. Figure 9 shows H${\alpha}$ and red continuum surface
brightness profiles for the same parameters as in Figure 8 but for
the metallicity of gas Z$_{\odot}/5$, which gives more satisfactory
fit to observational data.

The surface brightness profiles were then used to
calculate the luminosity of pre-collisional stellar disk of each radial zone.
The total luminosity of each zone is obtained by summing the luminosities
of pre- and post- collisional stars, which were then used to calculate the
color profiles.
The pre-collisional stellar disk seems to lack any bulge, and hence, 
the Cartwheel was most probably a late-type spiral (Higdon 1996).
We assume that the pre-existing old disk of the Cartwheel has color 
properties typical to those in the late-type disk galaxies. 
As an example we used de Jong's (1996) color profiles for the Sbc
galaxy UGC\,01305.  
The resulting color profiles are plotted in Figures 10--11 for
Z$_{\odot}/20$ and Z$_{\odot}/5$ metallicities respectively.
The filled triangles in the Figures indicate the extinction corrected 
observed colors of the nine rings in the Cartwheel's disk, and its nucleus.
The colors of galaxy UGC\,01305 are marked by open circles. 
The resulting colors of the mixed stellar populations are shown by the open
squares. The outer three radial zones, 9, 8 and 7 are identified both for the
observed and model points.
It can be seen that the observed colors are best reproduced in
the model with exponentially decreasing gas surface density distribution
and with metalicity of gas Z$_{\odot}/20$ (Figure 10). This model, however,
does not reproduce well the red continuum surface brightness profile.
The model with the metallicity of the star forming gas Z$_{\odot}/5$
satisfactory reproduces the range of the observed colors and their spatial
sequence.  
This is not the case if the metallicity of the stars formed in the wave
was solar, as illustrated in Figure 12. All the notations of the
previous figure are retained, and hence the open squares represent the
mixed model colors. This model neither explains the range of observed
colors nor the spatial sequence of the colors.

\section{SUMMARY AND DISCUSSION}

This paper has focused on the optical properties of disk galaxies
with large-scale star forming rings. We considered an improved 
model of a star forming wave, and used the population
synthesis approach to calculate the disk color gradients produced by an
outwardly propagating wave of star formation. The results of our 
calculations can be summarized as follows:

1. Sub-solar metallicities (at least by a factor of 2.5) are  
necessary for generating color profiles which redden systematically
towards the center.
This result is consistent with the observed metallicity 
in the Cartwheel, where such color gradients are observed.

2. The theoretical colors of the stellar populations born in the
wave propagating in a purely gaseous disk are much bluer than the 
observed colors of the Cartwheel's disk. 
A reddening gradient caused by interstellar dust cannot explain the
discrepancy.
Instead, an explanation of the Cartwheel's colors requires the presence of old 
stellar disk extending out to the present position of the ring.
The existence of a pre-collisional nucleus in the Cartwheel has been known for
more than two decades, but this is the first unambiguous evidence for a 
large stellar disk in the Cartwheel.

Simple models used in this paper leave aside many important aspects of 
the dynamics of ring galaxies. They  allow, however, to make a conclusion
on the nature of pre-collisional Cartwheel galaxy.
Higdon (1996)
assumed that a pre-collisional Cartwheel would have appeared as a
small spiral embedded in an extensive low surface density HI disk.
Our results show that the old stellar populations must exist in the entire disk
of the Cartwheel in order to understand the observed color gradients.

These conclusions raise the question of the survival of
the pre-collisional disk structure during an encounter. 
Athanassoula, Puerari \& Bosma (1997) numerically studied collisions
of intruders with barred disk galaxies. They found that an intruder with
mass equal to ten percent of the mass of a target barred galaxy leaves
the bar un-destroyed. To our knowledge, there is no study of the
collisional disruption of disk galaxies that have well-developed spiral 
structure. The stability properties of gravitating
disks can be drastically changed by a few percent admixture of cold gas.
If the pre-collisional disk of Cartwheel was gas rich, and most of the gas
was burnt in the star-forming wave or stripped from the galactic plane,
any pre-collisional spiral structure would not have survived.
All these questions remain for future investigations.


\begin{acknowledgments}
The authors thank 
an anonymous referee for his useful comments,
W. Wall for reading the manuscript which resulted in improved presentation,
and Roelof de Jong for providing his observational data in
electronic form. VK acknowledges Prof. S. Miyama for his hospitality, 
and the National Astronomical
Observatory of Tokyo for providing COE fellowship.
This work has been partly supported by CONACyT research grant
211290-5-25869E.

\end{acknowledgments}
\newpage
%

%
\clearpage


\begin{figure}[htb]
\centerline{\psfig{figure=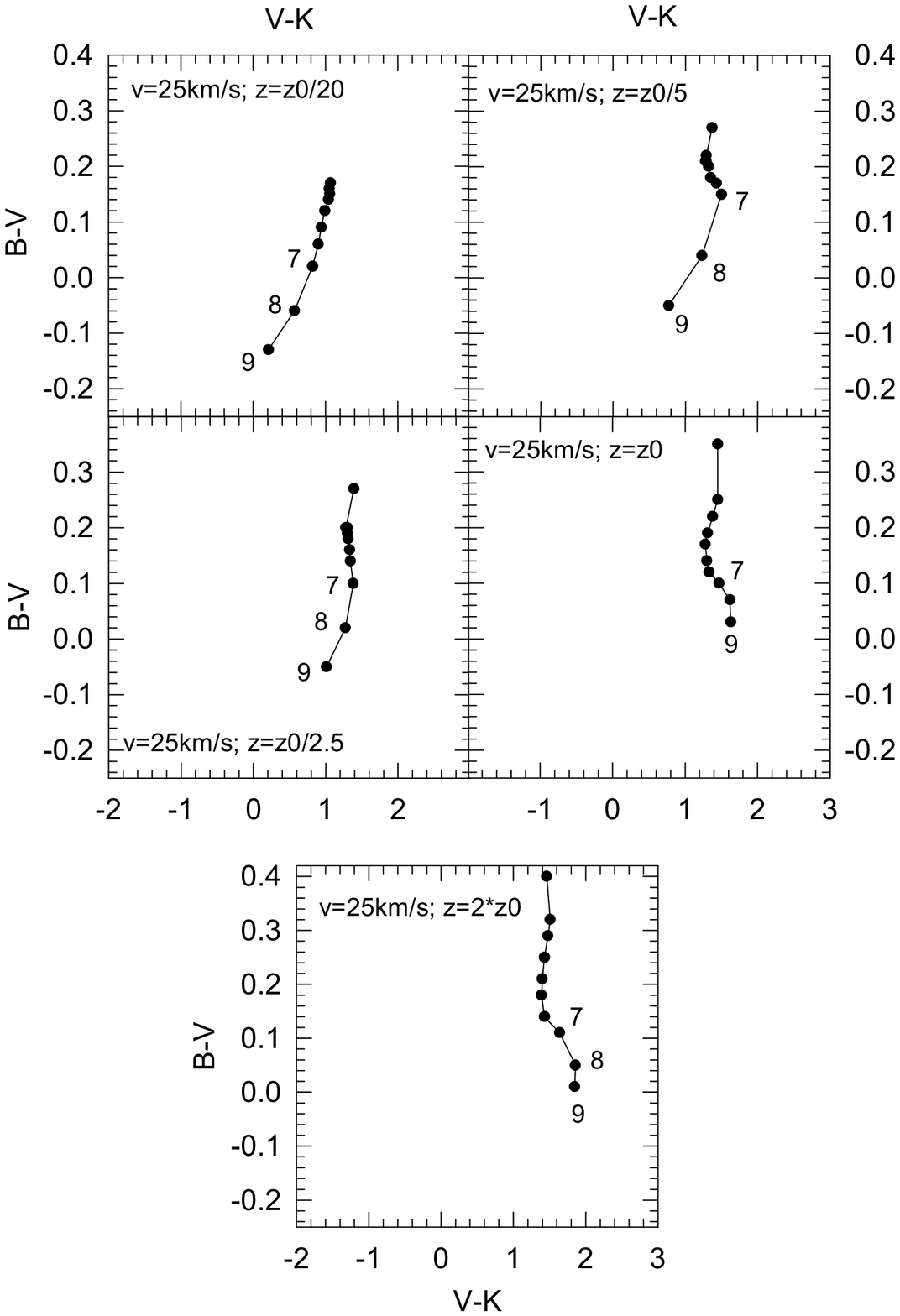,width=11cm}}
\figcaption[cw-col-figure1.ps]
{The $B-V$, $V-K$ color plots computed for nine
annular zones of a model ring galaxy.
The star-forming density wave propagates with a velocity 25 km\,s$^{-1}$
in a purely homogeneous gaseous disk. 
The plot illustrates the radial dependence of
colors within the disk for five different metallicities of the star-forming gas
Z$_{\odot}/20$, Z$_{\odot}/5$, Z$_{\odot}/2.5$, Z$_{\odot}$, and 
$2 Z_{\odot}$.
Colors have regular reddening towards the disk center if the initial
metallicity of gas is low.}
\end{figure}

\begin{figure}[htb]
\centerline{\psfig{figure=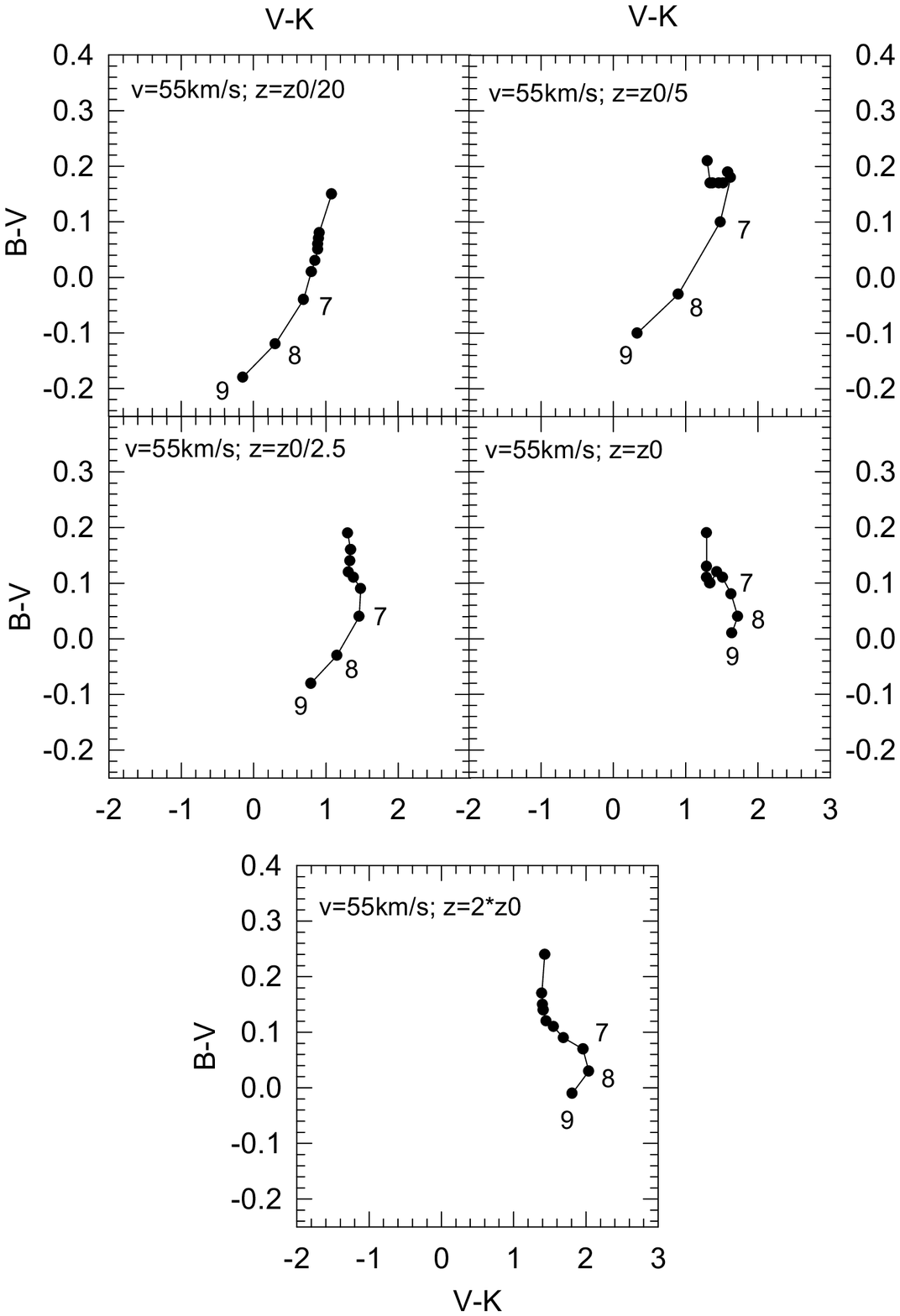,width=11cm}}
\figcaption[cw-col-figure2.ps]
{Same as in Fig. 1, but for a wave velocity of v=55 km\,s$^{-1}$.}
\end{figure}

\begin{figure}[htb]
\centerline{\psfig{figure=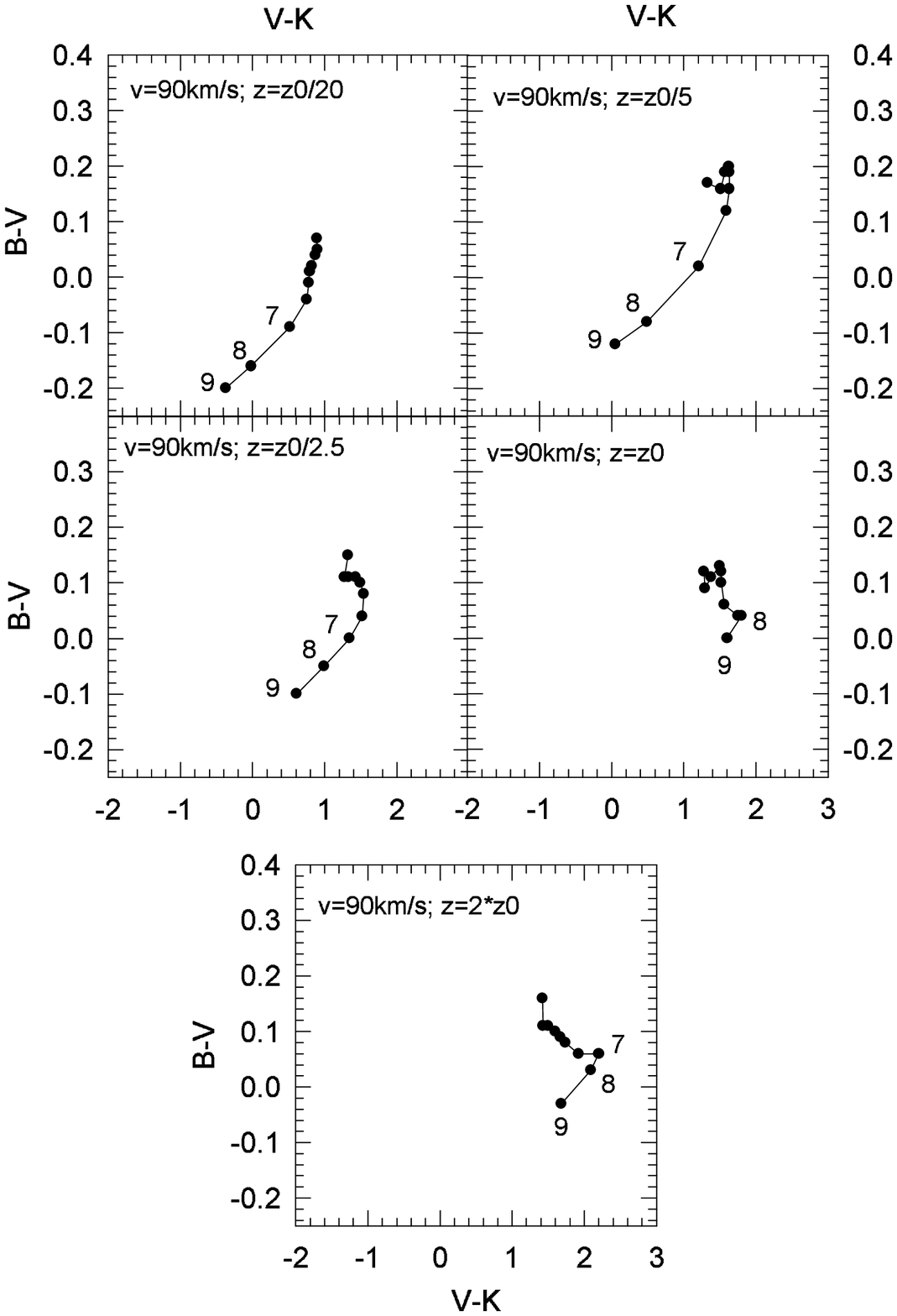,width=11cm}}
\figcaption[fig3.ps]{Same as in Fig. 1, but for a wave velocity of 90
km\,s$^{-1}$.}
\end{figure}

\begin{figure}[htb]
\centerline{\psfig{figure=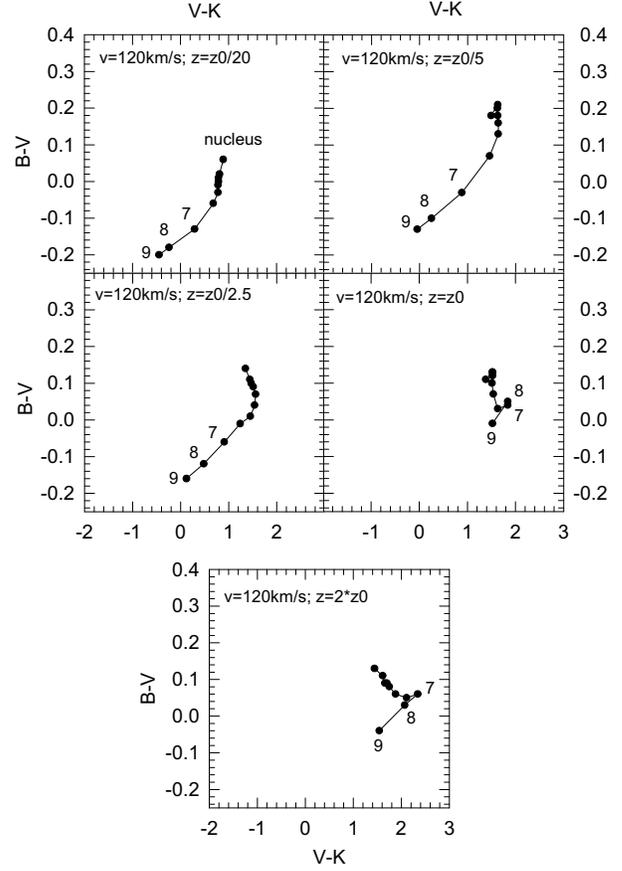,width=11cm}}
\figcaption[fig4.ps]{Same as in Fig. 1, but for a fast star-forming 
wave (v=120 km\,s$^{-1}$).}
\end{figure}

\begin{figure}[htb]
\centerline{\psfig{figure=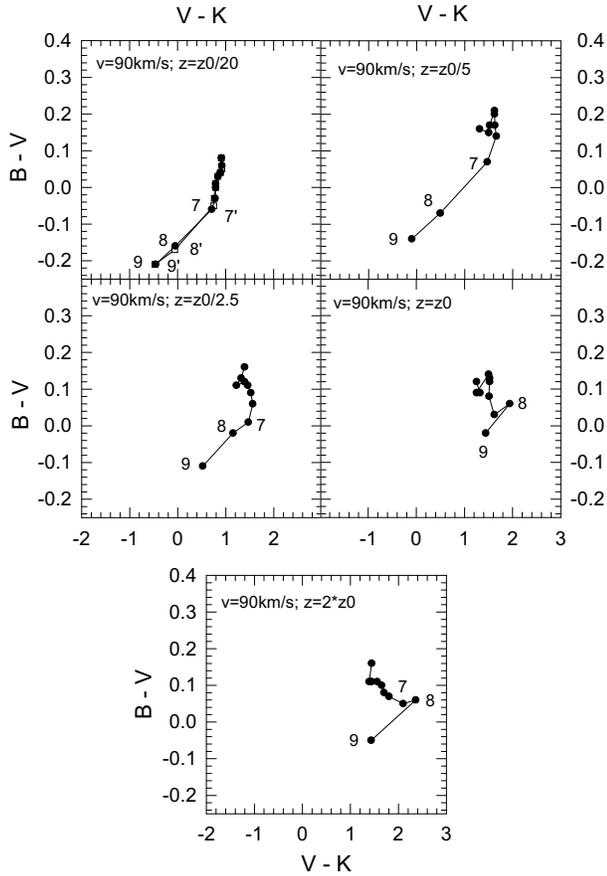,width=11cm}}
\figcaption[fig5.ps]{The $B-V$, $V-K$ color plots computed for the FIF wave
propagating with a velocity 90 km\,s$^{-1}$. The color diagrams for the FIF
wave are similar to the color diagrams of the density wave propagating with
the same velocity (Fig 3). The top-left frame shows also color diagram 
computed for the FIF wave propagating in the disk with an exponential 
surface density distribution (open squares) used for the modeling of the 
color properties of the Cartwheel galaxy.}
\end{figure}

\begin{figure}[htb]
\centerline{\psfig{figure=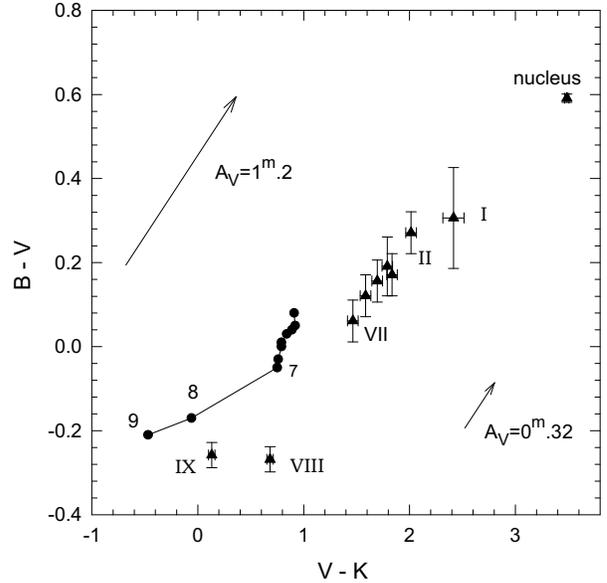,width=11cm}}
\figcaption[fig6.ps]{The model ring galaxy colors compared with the
observed colors of the Cartwheel. The large filled triangles 
mark the $B-V$, $V-K$ colors observed in the Cartwheel galaxy (MAH).  
The colors of the two outer photometric rings, VIII and IX
were corrected for extinction using $A_v = 1^m.2$. The colors of 
the inner photometric rings and the Cartwheel's nucleus were corrected 
using $A_V = 0^m.32$. The filled circles show the $B-V$, $V-K$ colors produced
by the star-forming wave propagating in a purely gaseous (Z$_{\odot}/20$) disk
with the  velocity 90 km\,s$^{-1}$. The model colors 
are bluer than the observed colors in the Cartwheel.
The values of extinction correction are shown by the arrows.}
\end{figure}

\begin{figure}[htb]
\centerline{\psfig{figure=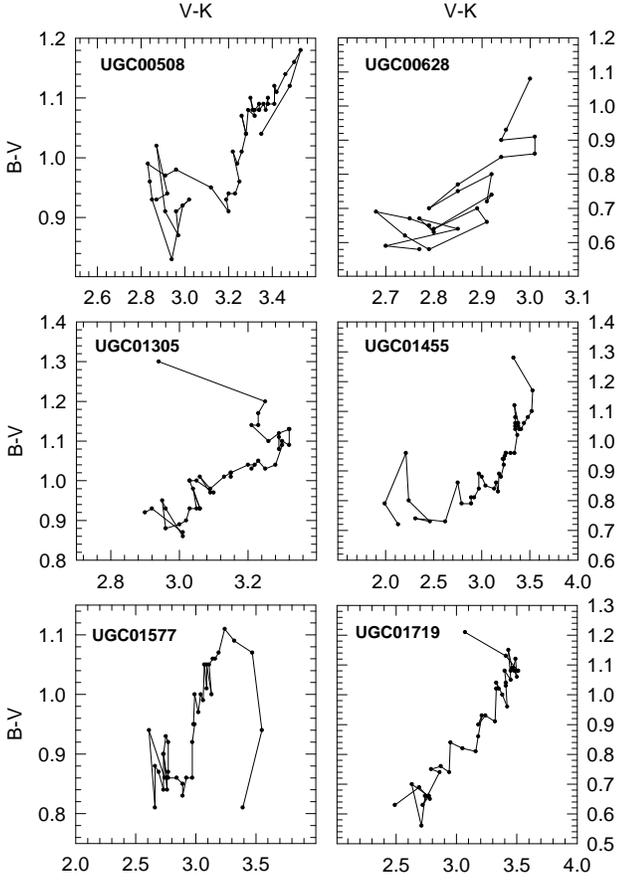,width=11cm}}
\figcaption[fig7.ps]{The color-color ($B-V$, $V-K$) plots of six disk
galaxies as measured by de Jong (1996).}
\end{figure}

\begin{figure}[htb]
\centerline{\psfig{figure=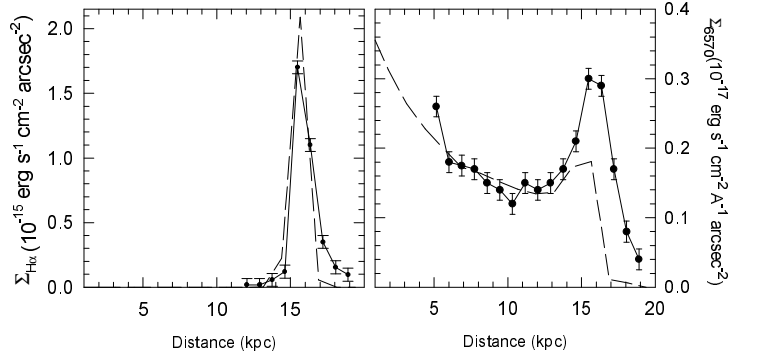,width=11cm}}
\figcaption[fig8.ps]{The red continuum and \ha\ surface brightness
distributions produced by the wave of star formation propagating
in a gaseous disk with exponential surface density profile with 
the parameters $A = 5\times 10^7$\msun\,kpc$^{-2}$, $H=11 kpc$\,
and gas metallicity Z$_{\odot}/20$. An underlying old stellar disk 
with $\mu_0(V)=23.5$~mag\,arcsec$^{-2}$ and $R_0=5$~kpc is added to the 
model ring galaxy.}
\end{figure}

\begin{figure}[htb]
\centerline{\psfig{figure=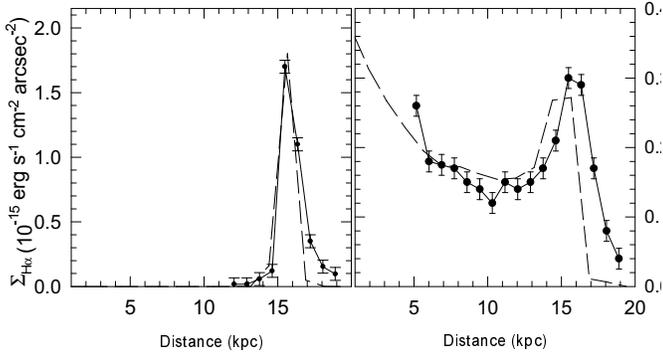,width=11cm}}
\figcaption[fig9.ps]{The red continuum and \ha\ surface brightnesses
in the wave with the same parameters employed in Figure 8 but with
metallicity Z$_{\odot}/5$.}
\end{figure}

\begin{figure}[htb]
\centerline{\psfig{figure=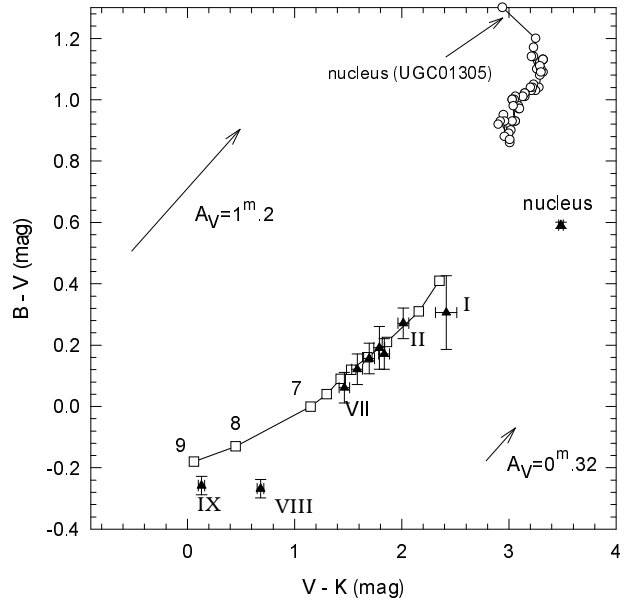,width=11cm}}
\figcaption[fig10.ps]{The model colors computed for the wave of star formation
propagating in a gaseous disk with the parameters employed in Figure 8.
An underlying disk added
to the model ring galaxy. The large filled triangles 
mark the extinction corrected colors of the Cartwheel galaxy. 
The open circles indicate the observed colors for galaxy  UGC\,01305.
The open squares show the colors obtained as a result
of mixing of the colors produced by the star-forming wave propagating 
in a purely gaseous (Z$_{\odot}/20$) exponential disk, 
with the observed colors in
UGC\,01305. The mixed colors are close to the observed ones in the Cartwheel.}
\end{figure}

\begin{figure}[htb]
\centerline{\psfig{figure=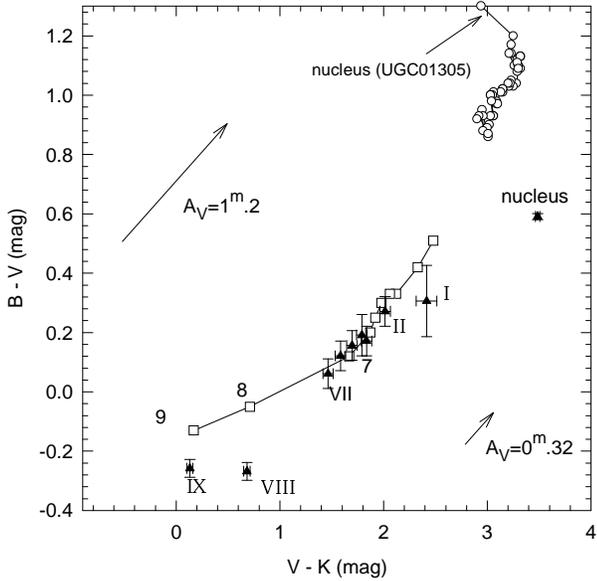,width=11cm}}
\figcaption[fig11.ps]{Same as in Fig. 10, but for the initial
metallicity of the gaseous disk Z$_{\odot}/5$.}
\end{figure}

\begin{figure}[htb]
\centerline{\psfig{figure=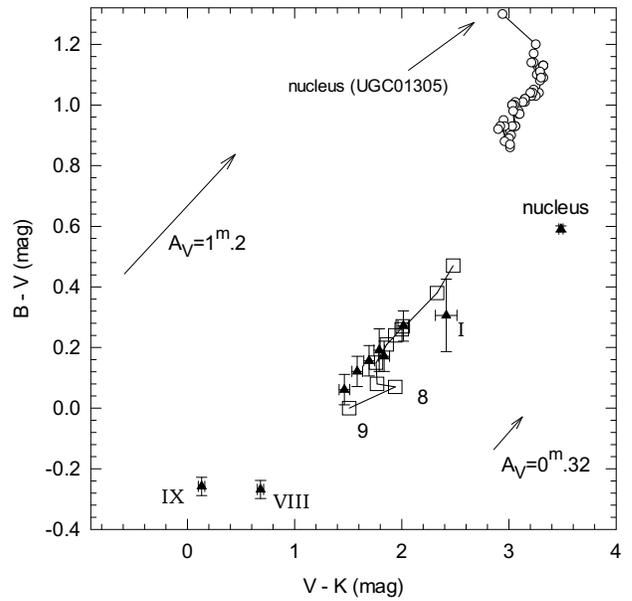,width=11cm}}
\figcaption[fig12.ps]{The model colors computed for the model
with the parameters employed in Figures 10 and 11, but for the initial
metallicity of the gaseous disk Z$=$\zsun. The observed colors are
in disagreement with the Cartwheel's observations, suggesting a
low metallicity for the star-forming gas in the Cartwheel.}                 
\end{figure}

\newpage

\begin{table}
\caption{Gas surface densities estimated from the
maximum allowed color excesses compared with the observed gas surface
densities.}
\vskip 1.5cm
\begin{tabular}{cccccc}
\hline
\hline
annulus
& $E(B-V)$  & $\Sigma_{HI+H_{2}}^{ext}$ &
$\Sigma_{HI}^{obs}$ & $\Sigma_{H_{2}}^{obs}$ &
$\Sigma_{HI+H_{2}}^{obs}$ \\ [-5pt]
  & mag &  $M_{\odot}/pc^2$  & $M_{\odot}/pc^2$ & $M_{\odot}/pc^2$ &
$M_{\odot}/pc^2$ \\
\hline
nucleus & 0.62 & 27.1 & $\le 0.3$ & 20--82 & 20--82 \\ [-5 pt]
I & 0.37 & 15.9 & $\le 0.3$ &  20--82 & 20--82  \\ [-5pt]
II & 0.34 & 14.8 &  $ 4.5$ & ... & $ 4.5$ \\ [-5pt]
III & 0.27 & 11.8 & $ 4.5$ & ... & $ 4.5$ \\ [-5pt]
IV & 0.27 & 11.8 & $ 4.5$ &  ... & $ 4.5$ \\ [-5pt]
V & 0.27 & 11.6 & $ 4.5$ &  ... & $ 4.5$ \\ [-5pt]
VI & 0.26 & 11.4 & $ 4.5$ & ... & $ 4.5$ \\ [-5pt]
VII & 0.23 & 10.1 & $  4.5$ & ... & $ 4.5$ \\ [-5pt]
VIII & 0.31 & 13.6 & $ 16.5$ & ... & $ 16.5$ \\ [-5pt]
IX & 0.36 & 15.7 & $ 16.5 $ & ... & $ 16.5$ \\
\hline
\end{tabular}
\end{table}

\end{document}